\documentclass[pre,twocolumn,floatfix]{revtex4}

\usepackage{graphicx}

\usepackage{wasysym}
\usepackage{color}
\newcommand{\RR}[1]{{\color{black} #1}}

\begin{document}

\title{Magnetic Resonance Imaging of Granular Materials}

\author{Ralf Stannarius}
\affiliation{Institute of Experimental Physics, Otto-von-Guericke-University, Universit\"atsplatz 2, D-39106 Magdeburg, Germany}

\date{\today}

\begin{abstract}

Magnetic Resonance Imaging (MRI) has become one of the most important tools to screen humans in medicine,
virtually every modern hospital is equipped with an NMR tomograph. The potential of NMR in
3D imaging tasks is by far greater, but there is only 'a handful' of MRI studies of particulate matter.
The method is expensive, time-consuming, and requires a deep understanding of pulse sequences,
signal acquisition and processing.
We give a short introduction into the physical principles of this imaging technique, describe its
advantages and limitations for the screening of granular matter and present a number of examples of
different application purposes, from the exploration of granular packing, via the detection of flow
and particle diffusion, to real dynamic measurements.
Probably, X-ray computed tomography is preferable in most applications, but fast imaging of single
slices with modern MRI techniques is unmatched, and the additional opportunity to retrieve spatially
resolved flow and diffusion profiles without particle tracking is a unique feature.
\end{abstract}

\maketitle

\section{Introduction}

Nuclear Magnetic Resonance (NMR) imaging is a mature but still rapidly developing technique that has
an ample range of clinical applications today. It is impressive to recall the history of this imaging
method during the past 80 years:
The first observation of the absorption of electromagnetic waves by nuclear spins in a molecular beam  was reported in 1937 by Isidor Isaac Rabi (1898-1988), who modified the classical Stern-Gerlach experiment by adding an oscillating magnetic field \cite{Rabi1938}.
In 1944, he received the Nobel Prize for this discovery. The method soon proved to be very useful for
the study of magnetic and mechanic properties of nuclei. The precise measurement of resonance
frequencies of nuclei laid the foundation for countless applications, from atomic and molecular beam experiments to Masers and contemporary NMR imaging.
Edward Mills Purcell (1912-1997, Nobel Prize 1952) and Felix Bloch (1905-1983, Nobel Prize 1952) independently discovered the magnetic resonance phenomenon in condensed matter in 1946 \cite{Purcell1946,Bloch1946}. Progress in the field grew steadily, Herman Y. Carr (1924-2008) was the first to apply defined magnetic field gradients to achieve a (one-dimensional) spatial mapping of
nuclear magnetic spins.
This discovery would probably not have had the giant impact we state today, had not
Raymond Damadian (born 1936) demonstrated that tissues and tumors can be distinguished by the
relaxation times of their nuclear spins \cite{Damadian1971}.
Medical application boosted the development of the method.
Paul Lauterbur (1929-2007, Nobel Prize 2003) and Sir Peter Mansfield (born 1933, Nobel Prize 2003) substantially contributed to the development of Magnetic Resonance Imaging (MRI), the former presented the first MR images of a living mouse \cite{Lauterbur1974},
the latter achieved immense progress in image reconstruction procedures, multipulse NMR,
and the mathematical analysis of the radio signals.
Efforts to establish MRI for clinical applications were further triggered by Hounsfield's introduction
of X-ray based computed tomography \cite{Hounsfield1973} in 1973 as a competing very successful imaging method.
Hospitals soon became interested to spend large sums for medical imaging, where X-ray Computed Tomography (X-ray CT) and MRI became strong competitors.

Today, contemporary MRI equipment achieves remarkable benchmarks. Structural MRI nowadays has the
sensitivity to detect microliter volumes. Special
techniques can also encode local velocities in the recorded signals and map local fluid flow.
Diffusion tensor imaging (DTI) can quantify diffusion in fluids and thereby motion on a scale much below the size of a voxel (volume element of the tomogram).
Spatially resolved diffusion tensors can be reconstructed.
Functional MRI (fMRI), as another example, allows to examine brain activities from blood flow in the cortex, this is commonly achieved with blood-oxygen-level dependent (BOLD) contrast measurements.

Many of these highly sophisticated methods have the potential to explore ensembles of granular particles as well. Among the interesting problems for structural NMR imaging is the revelation of packing structures,
segregation, orientational order of grains and the composition of granular mixtures.
Flow measurements with rheo-NMR \cite{Callaghan2006} can be useful in slurries or fluids inside porous structures. Furthermore, \RR{grain displacement} caused by agitation can be mapped.

Before we describe applications in particle imaging, it will be instructive to introduce the basic physics of NMR and principles of MR Imaging.
The fundamentals of NMR are found in quantum mechanics. However, for most of
the techniques described in this overview, only macroscopic physical quantities of
large spin ensembles are relevant. We can therefore resort \RR{to} classical, phenomenological models
and get around the quantum mechanical description.
Focus is laid on information needed for applications in grainy matter and particle imaging.
The introduction shall help the reader to decide whether MRI may be useful
to solve a certain particle imaging problem, and which technique may give optimal results. We will
then discuss typical applications in granular matter research, and finally compare advantages and drawbacks of MRI to competitive techniques.

\section{Basics of NMR and MRI}

\subsection{Nuclear spins, nuclei suitable for NMR}

The origin of the electromagnetic signal in NMR are nuclear spins, which are related to
a very small magnetic moment. The spin of a nucleus is composed of the spins of the individual nucleons.
The simplest atom, $^1$H, with only one proton, has the spin quantum number $I=1/2$.
The $^1$H nucleus has two energy levels,
$E = - \gamma I_z  B=\pm \hbar \gamma B/2$, with the gyromagnetic ratio $\gamma$.
Transitions between these two levels are related to emission/absorption at the Larmor frequency
\begin{equation}
\label{eq:larmor}
\nu_L = \frac{\omega_L}{2\pi}= \frac{\gamma}{2\pi} B.
\end{equation}
For the proton, $\gamma=2\pi\cdot 42.58$~MHz/T.
In nuclei consisting of multiple nucleons, the individual spins may pairwise compensate each other,
so that not all isotopes possess a magnetic moment. 
By far most frequently used for NMR experiments is the hydrogen $^1$H nucleus. Table \ref{Tab1} lists some of the
most common isotopes for NMR spectroscopy.
Nuclei with \RR{both} even proton numbers and neutron numbers (like the naturally highly abundant $^{12}$C
or $^4$He) have spin zero and are NMR-'invisible'.

\begin{table}[htbp]
\centering
{
		\renewcommand{\arraystretch}{1.5}
		\begin{tabular}  {l |c |c |r }
			
			\hline\hline
			  Isotope& Spin    &   Relative    &  NMR frequency [MHz]        \\
                     &         &   abundance [\%]  & (at 2.349 T)\\
			\hline
			$^1$H & 1/2  &  99.98   & 100.000     \\
			
			$^2$H & 1   &  1.5$\cdot 10^{-2}$   & 15.35     \\
			
			$^3$H & 1/2   &  0   & 106.67         \\
			
			$^3$He & \RR{3/2}  &  1.3$\cdot 10^{-4}$   & 76.18     \\
			
            $^{13}$C & 1/2 &  1.108   & 25.15     \\
			
            $^{14}$N &1 &  99.63   & 7.227      \\
			
            $^{23}$Na & 3/2 &  100   & 25.13        \\
			
            $^{27}$Al & 5/2 &  100   & 26.06      \\
			
            $^{29}$Si & 1/2 &  4.7   & 19.87      \\
			
            $^{31}$P & 1/2 &  100   & 40.49        \\

            $^{129}$Xe & 1/2 & 26.4 & 27.81\\
			\hline

		\end{tabular}
	}
	\vspace{5mm}
	\hspace{6cm}
	\caption{Table of natural abundances and Larmor frequencies of a selection of typical isotopes.
Because most of these isotopes have lower resonance frequencies than the $^1$H isotope,
$^1$H is almost exclusively the nucleus of choice in MR imaging.
}
	\label{Tab1}%
\end{table}

Medical MRI almost exclusively uses the $^1$H isotope. It is abundant in living bodies,
mostly in water and fat molecules. In non-medical particle imaging, one might in principle choose other
nuclei, but tomographs are usually tuned to the proton resonance frequency and other isotopes represent
nonstandard applications. Even though they may be used in principle, all published MRI studies in granular matter physics (except a few applications of hyperpolarized Noble gases mentioned below) have employed $^1$H resonance.

\subsection{Sensitivity}

The major problem of NMR signal detection is the very low population difference between the Zeeman levels, because of their low energy
difference, e.~g. $\Delta E \approx 7\cdot 10^{-6}kT$ at room temperature for protons at fields of the order of one Tesla. Thus the difference in the occupation of the Zeeman levels is only a handful of spins in one million. We mention in passing that
hyperpolarization of Noble gases like $^{129}$Xe or $^3$He by optical pumping (see, e. g. Ref. \cite{hyperpolarization}) can serve as a sophisticated but effective means to achieve huge excess populations (up to $70\%$) of one level, which can increase the NMR signal by several orders of magnitude. Such techniques have been successfully applied for dynamic flow measurements in granular matter
\cite{Wang2005}.

In the following, we will focus on the magnetization $\vec M$, which in equilibrium
is the sum of the nuclear magnetic moments $\vec\mu= \gamma I_z \vec e_z$ of the nuclei.
The equilibrium value $M_0$ in a magnetic field $B_0$ is
\begin{equation}
M_0=\frac{N \gamma^2\hbar^2I(I+1)}{3kT} B_0.
\end{equation}
Here, $k$ is the Boltzmann constant, $T$ the temperature of the spin system, N the number of spins.
Higher magnetic fields provide higher sensitivities because of the increased population density difference
\RR{(proportional to $B_0$) and higher energy of the radiation (linear with $B_0$). The noise level
also increases (approximately  with  ${B_0}^{1/2}$) because of the larger bandwidth in signal detection,
thus the signal to noise ratio (SNR) improves
only with a lower power than $B_0^2$.} Often, the SNR is assumed proportional to  $B_0^{3/2}$.
The conclusion is that for sensitivity reasons, it is generally advantageous to use high $B_0$ fields.
There are a few exceptions. For example, in grainy materials with large inhomogeneities of the magnetic
susceptibility it can be useful to reduce the involved statistical field gradients by choosing
lower $B_0$.
Today, typical NMR magnets provide fields from about 1 T to 23.5 T (the latter corresponding to a Larmor frequency of $\approx 1$~GHz).
Contemporary human MR scanners work at $B_0$ fields up to 9.4~T (400~MHz), and devices with up to 11.7~T
are under installation. Small animal scanners are in operation with fields as large as 17~T.

\subsection{Signal detection and processing, spectral shapes}

The simplest way to detect the NMR signal is the measurement of the absorption or emission of the sample during
the transition between the Zeeman levels, but such {\em continuous wave} spectrometers are nowadays rare,
they work slowly, since they require the scan of the complete interesting frequency range to obtain a spectrum. They are not suitable for MRI.

When the sample magnetization is not aligned parallel to $\vec B$, it precesses around the field direction according to the Bloch equation
\begin{equation}
\frac{d\vec M}{dt} = \gamma \vec M\times\vec B.
\label{eq:precess}
\end{equation}
This allows to employ a much simpler method to obtain the complete spectrum from the signal emitted by the precessing magnetization, with {\em pulsed} spectrometers. The magnetization is flipped out of equilibrium by a lateral rf pulse of appropriate frequency, field strength and duration. Thereafter, the Free Induction Decay (FID) signal is recorded and Fourier transformed to obtain the full spectrum.
The information obtained by both techniques is equivalent.

High resolution NMR of liquids can resolve very sharp lines, where signals from different chemical sites in a molecule can be distinguished (the local $B_0$ field is partially screened by electrons in chemical bonds).
The chemical shift is of the order of a few ppm, i. e. the local magnetic field is modified in the order of several $\mu$T due to the chemical environment.
In solids, the {\em dipolar} interactions between nuclear spins are not averaged by random thermal
motion, unless special techniques (Magic Angle Spinning and rf decoupling) are employed.
These dipolar fields broaden the spectra substantially.
The consequence is that intensities in spectra of solids are smeared out over a broad frequency range so
that the peak heights are orders of magnitude smaller. Common MR Imaging is practically impossible with  satisfactory resolution. Special techniques that have been proposed \cite{Samoilenko1988}
suffer from severe limitations. We thus register that
(with the exception of hyperpolarized gases mentioned above) only liquids are suitable for imaging.
MRI in particulate matter explores the distribution of liquids in the sample.
Particle imaging can be achieved either directly (the particles contain free liquid or are coated with a liquid layer) or indirectly (solid particles are embedded in a liquid and a 'negative' image is recorded).

\subsection{Pulsed NMR, relaxation, FID}

As mentioned above, the most efficient way to obtain the spectrum is to force the magnetization into a
plane perpendicular to the $\vec B_0$, where it precesses with the Larmor frequency $\omega_L$. This is achieved by a radio
frequency (rf) pulse  $\vec B_1=(B_1 \cos\omega_L t,B_1 \sin\omega_L t,0)$ at the same frequency.

\begin{figure}[htbp]
\includegraphics[width=0.8\columnwidth]{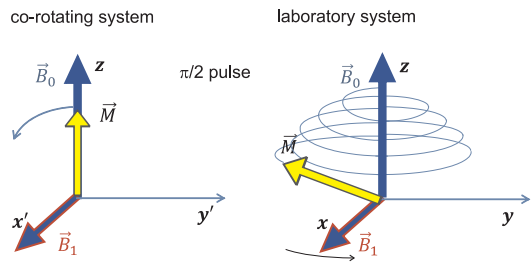}
	\caption{$\pi/2$ rf pulse applied to flip the magnetization $\vec M$ into a plane perpendicular to the
$\vec B_0$ field, $\vec M$ is rotated into an orientation perpendicular to $\vec B_0$.}
\label{fig:pulse}
\end{figure}

In the co-rotating frame, this $B_1$ field is stationary, the
magnetization thus rotates around $\vec B_1$. The duration and magnitude of the rf pulse are adjusted such that $\vec M$ performs a quarter rotation ($\pi/2$ pulse. Figure \ref{fig:pulse} sketches the magnetization dynamics
in the co-rotating $(x',y',z)$ and laboratory $(x,y,z)$ frames.
\RR{Technically, one can use an oscillating linear $B_1$ field which contains two components rotating with and
against the precession of $\vec M$, the counter-rotating component averages out completely.}

\begin{figure}[htbp]
\includegraphics[width=0.95\columnwidth]{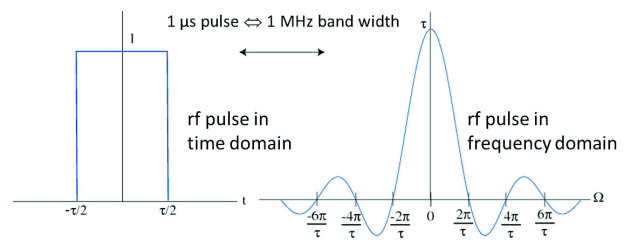}
	\caption{Exemplary plot of a rectangular $\pi/2$ pulse of duration $\tau$ in the frequency and time domains.
Narrow frequency bands are selected with long $\tau$ and, vice versa, broad band excitation with short $\tau$.}
\label{fig:bandwidth}
\end{figure}

In the frequency domain, an rf pulse with finite duration $\tau$ has a characteristic frequency spectrum that
guarantees that not only the spins at the Larmor frequency, but also those in an adjacent frequency band are
excited (Fig.~\ref{fig:bandwidth}). Keeping the flip angle $\propto B_1\tau$ constant, one can control the width of the excited frequency band by proper choice of $\tau$ (see Sec.~\ref{sec:imagingTechniques}).

After the excitation pulse, the magnetization gradually relaxes back to the equilibrium value by $T_2$ relaxation (spin-spin
interactions, de-focusing of the transverse magnetization) and $T_1$ relaxation (spin-lattice interactions, re-establishing of the longitudinal component). Equation~(\ref{eq:precess}) is modified by two corresponding terms,
\begin{equation}
\frac{d\vec M}{dt} = \gamma \vec M\times\vec B - \frac{M_x\vec e_x+M_y\vec e_y}{T_2}
- \frac{(M_z-M_0)\vec e_z}{T_1}
\label{eq:precess2}
\end{equation}

\begin{figure}[htbp]
\includegraphics[width=0.95\columnwidth]{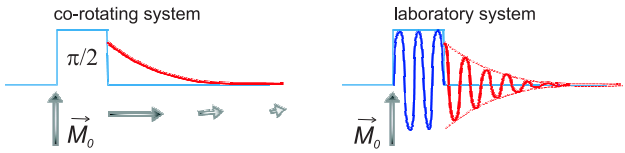}
	\caption{Evolution of the transverse magnetization after a $\pi/2$ pulse in the co-rotating (left) and laboratory (right) frames. The blue graph sketches the rf excitation pulse, the red curves $M_{y'}$ and $M_y$, resp.}
\label{fig:FID}
\end{figure}

The evolution of the transverse magnetization is
 illustrated in Fig.~\ref{fig:FID}. It is important to note that only the precessing component
of the magnetization, perpendicular to the $\vec B_0$ field, contributes to the recorded signal.
Both $T_1$ and $T_2$ are characteristic parameters that can be used to
distinguish or to mark spins in different environments. Typical values e. g. for oxygenated blood are
200 ms for $T_2$ and around 1.5 s for $T_1$. Local magnetic field inhomogeneities, caused for
example by inhomogeneous susceptibilities in granular matter, shorten $T_2$ to the effective value
$T_2'$, and they broaden NMR spectra.

\subsection{Echoes}

Rf pulses of appropriate amplitude and duration can flip the magnetization by arbitrary angles. For example, a $\pi$ pulse inverts $\vec M_0$
without leaving a transverse magnetization. It is useful for a direct measurement of $T_1$ by a $\pi$ pulse, followed
by a $\pi/2$ pulse with variable delay and subsequent recording of the FID (inversion recovery technique).
A similar procedure can be employed by labeling spins with respect to their $T_1$ \cite{Bydder}.

\begin{figure}[htbp]
\includegraphics[width=0.95\columnwidth]{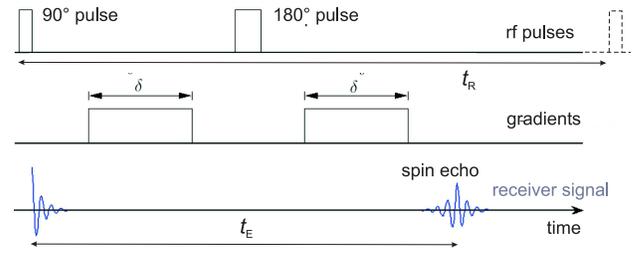}
	\caption{Pulse sequence for the simple Hahn echo (top graph),
it can serve (without the field gradients) as a simple
means to measure the true $T_2$ relaxation time. In combination with field gradients(middle graph),
one can use it to measure diffusion and drift. Relevant parameters are the echo time
$t_E$ and the repetition time $t_R$.}
\label{fig:spinecho}
\end{figure}

Following a $\pi/2$ pulse and FID, the $\pi$ pulse can be used to re-establish transverse magnetization by reversing defocusing processes of the sample magnetization due to magnetic field inhomogeneities
(Figure~\ref{fig:spinecho}). This is helpful in several aspects. First,
one can get another free induction signal, a so-called spin echo, after some delay. Second, one can determine the true $T_2$ by eliminating field inhomogeneity effects, and third, one can refocus the magnetization after deliberate defocusing with gradient fields. A linear magnetic field gradient $\vec G$ changes the local Larmor frequency of Eq. (\ref{eq:larmor}) to
\begin{equation}
\nu_L = \frac{\omega_L}{2\pi}= \frac{\gamma}{2\pi} (B_0 + \vec r \cdot \vec G).
\end{equation}
If the time between $\pi/2$ pulse and echo is short enough so that the spins did not move during
that interval, one can reverse field gradient effects completely. If the delay is sufficiently long so that
the spins have changed positions by diffusion or drift, the signal damping and/or phase shift
in a gradient experiment can serve as a tool to quantify diffusion or drift velocities.

\begin{figure}[htbp]
\includegraphics[width=0.95\columnwidth]{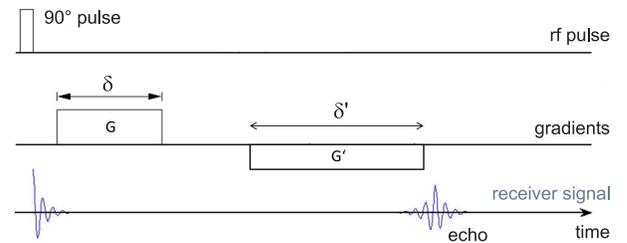}
	\caption{Pulse sequence for the gradient echo, it can serve as a tool to
determine diffusion and drift velocities. Gradient strength and duration must obey the relation
$G \delta = - G'\delta'$
for complete rephasing of \RR{spins of non-moving nuclei}.}
\label{fig:gradecho}
\end{figure}

\begin{figure}[htbp]
\includegraphics[width=0.85\columnwidth]{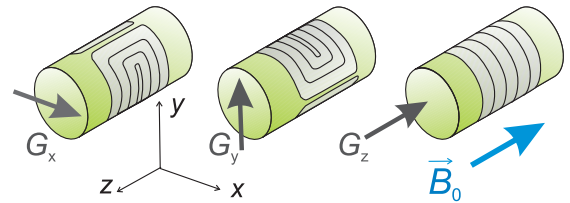}
	\caption{Realization of $x$, $y$, and $z$ gradient coils for MR imaging experiments.}
\label{fig:coils}
\end{figure}

Refocusing of the transverse magnetization after application of a magnetic field gradient can also be achieved by applying a gradient of opposite sign, provided that the spins have not moved between the start of the first and the end of the second gradient (Fig.~\ref{fig:gradecho}). Gradient echoes can thus be generated with a single $\pi/2$ pulse and two bipolar gradients that mutually compensate
each other (for non-relocating spins). By proper choice of three different gradient directions, the three
components of velocity fields or displacements by diffusion can be separately measured (Fig.~\ref{fig:coils}).

\section{MR Imaging techniques}
\subsection{Fundamentals}
\label{sec:imagingTechniques}
\begin{figure}[htbp]
\includegraphics[width=0.6\columnwidth]{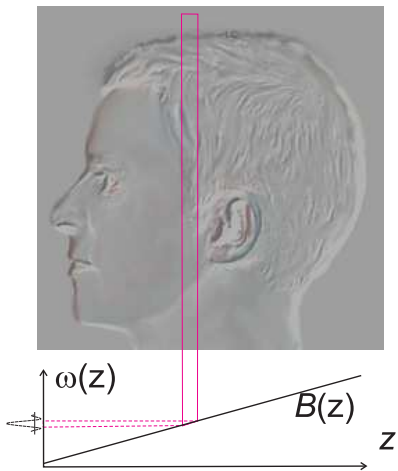}
	\caption{Selective excitation of one slice of the sample: The field gradient (here, in $z$ direction)
leads to an $z$-dependent resonance frequency, the $\pi/2$ pulse selectively excites only spins in a certain
$z$ interval.}
\label{fig:headslice}
\end{figure}

As indicated in the introductory section, field gradients provide a simple tool to label spins located
at a certain position by their frequencies. A linear gradient $G_z=dB/dz=\rm const$ can therefore be
used to map the spins along one coordinate by frequency. This can be exploited, in combination with
a narrow-band excitation
pulse, to excite only spins in a certain slice of a sample. Figure \ref{fig:headslice} depicts this
schematically. As shown in Fig.~\ref{fig:bandwidth}, a rectangular $\pi/2$ pulse has a broad frequency spectrum with side extrema. This is unwanted in slice selection where one attempts to have a rectangular characteristics of
the excitation in the frequency domain. Therefore, the $\pi/2$ pulse is preferentially shaped with an
appropriate envelope.
For example, a  $\sin (x)/x$ shaped rf pulse would have a rectangular shape in the frequency spectrum,
 $x=\omega_s t$ defines its spectral width $2\omega_s$.
Gaussian pulses are simpler to realize electronically, they have a Gauss-shaped frequency spectrum which
is the wider the shorter the pulse is.

Typical gradient fields of commercial human MR scanners are 0.02\dots 0.1~T/m. This can be 'translated' into a $^1$H
resonance shift of  1\dots 5 kHz/mm. Inhomogeneities of the diamagnetic susceptibility of the material may
distort the magnetic field locally and lead to artifacts in porous samples or near air interfaces. These local
field distortions are of the order of a few ppm. Consequently they may deteriorate the spatial resolution and
produce mapping errors in the submillimeter and even millimeter range. Paramagnetic grains in the sample  materials must be strictly avoided, since the distortion of the local field (of the order of hundreds of ppm) will most probably ruin the images. However, paramagnetic salts dissolved in embedding liquids
can be used as dopants to reduce the relaxation times.
It goes without saying that even small ferromagnetic particles will have catastrophic consequences in
MRI experiments.

Since the mapping is achieved via the Larmor frequency in a given magnetic field, it should be noted that this
technique strictly works only when all hydrogen nuclei in the sample have the same chemical environment.
If the imaged chemical is water, there is only one single line in the spectrum. In case more complex liquids
shall be imaged (e.g. oil contained in seeds in granular matter experiments),
typical widths of $< 5$ ppm $ \approx 1.5$~kHz of proton spectra may influence the resolution.
In that case, each line will contribute an image where the different chemical shifts manifest themselves in apparent spatial shifts,
producing shadows. In most cases, this problem can be ignored since the field gradients across one voxel are large enough compared to the chemical shift.
Of course, a chemical shift selective preparation pulse sequence in absence of gradients
can solve the problem \cite{Haase1985}, one can even image the distribution of different liquids separately.

With a single gradient, only one-dimensional (1D) images are achieved. In order to obtain 3D resolution, the position information is encoded sequentially with three orthogonal gradients in the following way:
\begin{enumerate}
\item {\em Layer selection gradient (z)}: During the $\pi/2$ and $\pi$ pulses, and only then, a gradient is applied in the first direction. As sketched in Fig.~\ref{fig:headslice}, this technique selects only spins in a certain slice of the sample.
    In all other sample regions, the equilibrium magnetization remains unchanged.
\item {\em Phase encoding gradient (y)}: Another gradient perpendicular to the first one is employed shortly after excitation gradient. This gradient encodes the $y$-position of the excited spins in a controlled way
    in the phase of the precession. In each line along $y$, the spins acquire a different phase depending on
    their $y$ coordinate.
    Phase encoding has to be repeated with different phase encoding strengths to uniquely encode the $y$-position.
\item {\em Read out gradient (x)}: During the acquisition of the signal, a third gradient is applied, perpendicular to the previous two. Spins in each column of the image acquire different precession rates and thus emit at different Larmor frequencies. This frequency encoding carries information on the $x$ position.
\end{enumerate}

\begin{figure}[htbp]
\includegraphics[width=\columnwidth]{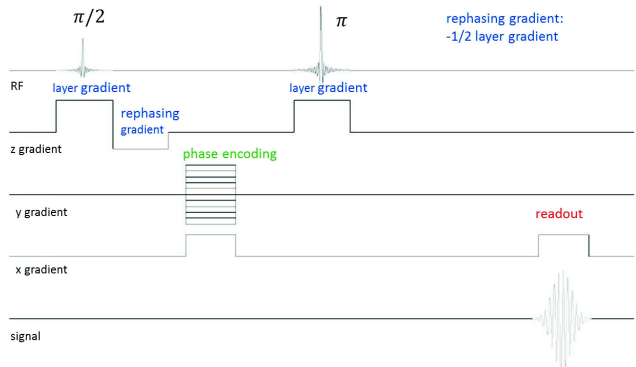}
	\caption{Schematic sketch of the sequence of rf pulses and gradients for a simple MR imaging experiment. The sequence has to be repeated with varied phase encoding gradients ($y$) in order to obtain
the $y$ information.
The evaluation of the acquired signal yields the spin density $\rho(x,y,z)$ in one $z$ slice of the sample.}
\label{fig:fullscheme}
\end{figure}

Schematically, this is shown in Fig.~\ref{fig:fullscheme}.
This procedure yields the complete 2D information about the distribution of spins in one slice of the sample.
This technique
is fast and can be used if one is interested only in the structure of one particular cross section. Without the layer selection gradient, one can excite the complete sample and one obtains a 2D image
averaged over all $z$ layers (comparable to an X-ray transmission image).

In order to construct the complete 3D image, it is necessary to repeat the complete procedure for each
slice in $z$ direction. A time consuming feature of the imaging process is
that normally one has to wait for a sufficiently long time period $t_R$ between subsequent $\pi/2$ rf pulses.
This repetition time $t_R$ should be of the order of $T_1$ to ensure that a substantial part of the equilibrium magnetization $M_0$ is reestablished before the next $\pi/2$ pulse is applied. However, there are workarounds that are described below.

\RR{A concept widely used to represent the raw data in MRI is the $k$-space \cite{Twieg1983}.
It comprises the complex 2D or 3D Fourier transform of the image.
In practice, the term often describes the temporary discrete image space after data acquisition.
The 2D $k$-space is in the spatial frequency domain with coordinates
 $   k_f =  \gamma G_y m \Delta t$   {(frequency encoded)} and
$    k_p = {\gamma }n\Delta G_z\tau$  {(phase encoded)}, $m,n$ are integers.
Typically, the number of rows and columns in $k$-space is the same as the final image.
If the transverse magnetization in the image is a real quantity, then the signal $S(k_f,k_p)$ is the
complex conjugate to $S(-k_f,-k_p)$, information is redundant.
}

\subsection{RARE and Echo-planar imaging (EPI)}

In conventional Spin Echo imaging, the repetition time $t_R$ between subsequent $\pi/2$ pulses is
determined by the longitudinal relaxation time $T_1$ (hundreds of milliseconds to more than one second). If within this period only one single echo is acquired after the echo time $t_E \ll t_R$,
then the time after the first echo is practically 'wasted'.
The use of multiple echoes as a technique to overcome long $T_1$ as the limiting factor for
acquisition rates was proposed already in the beginnings of the development of magnetic resonance
imaging by Mansfield \cite{EPI-Mansfield} in 1977.
One can apply more $\pi$ pulses after the first spin echo and record further echoes after each recovery pulse. This technique uses the repetition time very efficiently.
The number of echoes within the period $t_R$ is called the echo train length.
The multiple echoes acquired in one $\pi/2$ shot can be exploited to acquire multiple phase-encoding lines during each $t_R$ interval, by changing the phase encoding gradients between subsequent echoes.
Of course, one has to consider $T_2$ relaxation, which is often much faster than $T_1$ relaxation.
This method is commonly referred to as Fast Spin Echo (FSE) or Rapid Acquisition with Refocused Echoes (RARE) (in Ref. \cite{Henning}: \underline{R}apid \underline{A}cquisition with \underline{R}elaxation \underline{E}nhancement). Proper distribution of the phase-encoding steps minimizes artifacts \cite{Henning} and may even enhance resolution.

The so-called Echo Planar Imaging (EPI) allows to reconstruct a complete 2D MR image from a single FID decay very efficiently, in milliseconds. The idea is, again, to re-establish the magnetization needed for the next phase encoding scan with an echo following the previous signal acquisition sequence, up to 128
echoes can be achieved in a single train. Today, the commonly established definition of EPI includes any rapid gradient-echo or spin-echo sequences in which the complete phase-space is traversed in one ('single shot EPI') or a small number of ('multi-shot EPI') $\pi/2$ excitations.
In some sense, FSE with very long echo trains can be considered the multi spin-echo version of EPL.
In biological applications, but also in particle imaging, these techniques can help to remarkably reduce motion-related artifacts. Snapshots of dynamic processes become possible.


EPI has become one of the most efficient MR Imaging techniques, certainly the most preferred
for functional MRI, even though it is technically demanding and one pays for its advantages with
more or less severe artifacts. Usually, one has to condone with image distortions as the prize for
rapidity \cite{EPI}. In clinical applications, the high acoustic noise level is also not optimal.

\subsection{Ultra-fast imaging (FLASH)}
\label{sec:FLASH}

In Magnetic Resonance Imaging of dynamic processes, measurement time is often a decisive criterion.
As we have explained above, the main limiting factor in such experiments is the longitudinal relaxation time, $T_1$.
If one is interested only in the 1D profile along one direction in the sample, or a 2D image, then a single shot can be recorded within a few milliseconds, but the temporal resolution of successive images
is determined by the repetition rate of the experiment. After an initial $\pi/2$ rf pulse,
the magnetization along the $\vec B_0$ field is extinguished and one has to wait for a time period
of the order of $T_1$ for some longitudinal magnetization to be re-established.
This can be several hundred milliseconds up to seconds.

An elegant workaround that is used in most ultrafast MRI experiments in granular matter is the Fast Low Angle SHot (FLASH) sequence \cite{Haase1986}. The idea is to generate the initial transverse
magnetization with a low flip angle, e. g. with a $\pi/12$ rf pulse. The transverse magnetization is then only $\sin \pi/12 \approx 26$~\% of the equilibrium $M_0$ instead of 100 \% with a $\pi/2$ pulse, but a
longitudinal component of
$\cos \pi/12 \approx 96.5$~\% of $M_0$ remains for the next shot. This allows
rapid repetition rates that can be orders of magnitude faster than $T_1$, with a tolerable intensity
loss of about $ 74$~\%.
The next FID can be generated without potential problems with $T_2$ relaxation that occur in the
multi-echo techniques.

\subsection{MRI velocimetry}
\label{sec:velocimetry}
The measurement of flow or displacement of particles can be achieved in two different ways \cite{Fukushima1999,Gladden2013R}.
The first method is based upon spin tagging \cite{Axel}. Its principle can be described simplified
as follows:
In a preparation step, a long rf pulse rotates the magnetization in the sample, while a linear magnetic field
gradient modifies the Larmor frequencies of the spins along one spatial coordinate. After the preparation rf pulse,
the magnetization is modulated sinusoidally along one spatial coordinate, i.e. one starts the MRI experiment with slices labeled by their magnetization.
If no motion takes place in the evolution time between the preparation and the signal acquisition (echo),
the MR signal will reflect the spin densities modulated with the sinusoidal envelope. If the
labeled spins change their positions during this period, the labeled regions will be correspondingly distorted or shifted. One can easily map the displacement field in the direction normal to the slices.
The magnitude of the effect can be controlled by the choice of the evolution time. It is even
possible to imprint a 2D modulation of the initial magnetization (e. g. square grids), this allows to map
2D motion profiles. In applications with granular materials, one will usually prepare the initial
grain ensemble, then perform a certain manipulation like shearing or shaking, and record the image data
after the container and granulate are at rest again.

A second method is the use of bipolar gradients as described in Sec.~\ref{sec:velocimetry}.
Such gradients embedded in the conventional imaging sequence allow to map the spatial distribution
of velocity components \cite{Hanlon}. One records two images with different gradient magnitudes and/or durations,
then the phases of both images are subtracted. The flow velocity along the applied gradient pair
is encoded in this phase shift.
One may extend the method to apply flow-compensating echoes. They refocus the signal in case of linear motions.

\subsection{Imaging of specially weighted spin densities}

The simple imaging technique introduced in the first subsection has found many
variations during the last decades. For example, one can vary the delay $t_E$ between the first and second
rf pulses to produce $T_2$ weighted images, the measured spin density will then be weighted with the factor $\exp(-t_E/T_2)$.
One can likewise vary the repetition time $t_R$, to get $T_1$ weighted images. These will be scaled by a factor
$1-\exp(-t_R/T_1)$ in such an experiment. Additional magnetic field gradients placed before and after the
slice gradient can provide diffusion weighted images \cite{DWI}. In particle imaging, these techniques
are of little importance, since usually the relaxation times of the imaged liquids are rather uniform
in the different components of the samples.

\section{MRI of granular materials}

Interest in MRI studies arose in the granular community more than two decades ago.
Probably the first demonstration that liquid state MRI can be applied in fluid-embedded granulates
was given by Altobelli in 1991 \cite{Altobelli1991}. MR Imaging is applicable when the sample contains
a liquid whose spatial distribution or motion is mapped.
Since this limits the choice of materials considerably,
most of the experiments were performed using oil-containing seeds (e. g. mustard or poppy seed)
\cite{Ehrichs1995} or pharmaceutical pills
which contain oil and are therefore detectable by NMR, or with 'NMR-invisible' solid glass or plastic
beads that are immersed in a liquid like water or oil \cite{Baldwin1996}. The latter samples produce a
'negative' image of the particles. One can likewise visualize of voids in porous materials, by filling them with an NMR-active fluid.
In rare cases, porous grains were soaked in liquids to make them visible in nuclear magnetic resonance \cite{Sommier2001}. Interestingly, some seeds like rice grains, that are in the right size range for MRI, are hardly detectable because they possess only small water or oil content.
MRI can detect shapes in the millimeter and submillimeter ranges and may resolve submillimeter grains or structures.
Typical sample dimensions are several centimeters, but in human scanners it is possible to scan
volumes of several liters in volume. In stationary samples, one can achieve good S/N ratios and satisfactory sensitivities by accumulating the signal, condoning long acquisition times. In dynamic measurements where time restrictions are critical, one can reduce experiment time by restricting to
2D images of slices of the samples. Fast techniques like EPI may reduce
acquisition times considerably, so that dynamic processes in the millisecond range become accessible.
Often, distortions of the images related to the latter technique are less critical than in clinical applications where MRI images must be mapped to structures that were reconstructed by other techniques like X-ray CT.

The most interesting scientific problems in granular physics that can be tackled by MR imaging are\\
(a) characterization of packing structures of grains in 3D containers which are not accessible optically,\\
(b) mixing and segregation of grains in granular ensembles under mechanical agitation such as shaking, shearing, rotating, or other,\\
(c) diffusion of grains or interstitial fluids, particle displacements and flow in granular or porous
systems, and\\
(d) the reconstruction of complex dynamic processes.\\
We will describe these aspects in the following subsections.

A review on the application of MRI to complex fluids including, e. g. drainage of slurries
has been given by Bonn et al.~\cite{Bonn2008}. In Ref.~\cite{Gladden2013}, one can find a helpful literature list of NMR experiments in reaction engineering, including some reference to MRI. The
special field of MRI in heterogeneous catalysis, which bears some connections to particle imaging problems, has been reviewed by Koptyuk \cite{Koptyuk2014} recently.

\subsection{Packing}

Packing problems in granular matter represent one of the oldest and yet still a very topical field of
research. The Kepler conjecture about the most efficient way to pack identical spheres was uttered
already hundreds of years ago (1611), but it remained unsolved until recently \cite{Hales}.
This problem is probably among the simplest questions to be solved in this area.
When one incorporates spatial
restrictions in one or two dimensions, or chooses slightly polydisperse, or ellipsoidal \cite{Donev}, or even less regularly shaped particles, the situation soon becomes very complex. While regular packing may be interesting primarily from a fundamental point of view, one often deals with irregular arrangements in practical situations.
For these types of packing, even of identical particles, the interesting questions involve the
characterization of a
random close packed state, i. e. the densest non-crystalline state, poured random packing, i. e.
packing structures that are adopted without compaction by shaking or other agitation, and the least dense
regular packing. The first experimental studies that dealt with such structures in the last century
had to rely on tedious destructive methods: the packing state was fixed first with paraffin or black
paint \cite{Bernal,Scott}, then the structure had to be disassembled sphere by sphere.
Tomograms can nowadays help to study local arrangements, particle contacts, packing density fluctuations,
crystalline regions and other features inside optically non-accessible bulk material.

 \begin{figure}[htbp]
\includegraphics[width=\columnwidth]{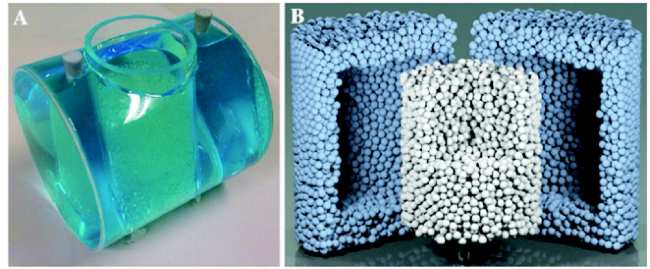}
	\caption{The image shows a) the phantom used for the MRI measurements and b) an example of a 3D
reconstructed packing. The outer spheres close to the container walls (shown in light blue) were discarded
in order to eliminate influences of the boundaries on the results.
Reproduced with permission from Appl. Magn. Reson. 46, 633 (2015). Copyright 2015 Springer Verlag.}
 \label{fig:balzan}
\end{figure}

A typical task that can be solved by tomographic imaging is the characterization of random packings of
identical spheres. This can be achieved, for example, by means of X-Ray CT \cite{Aste,Schaller}.
Balzan et al. \cite{Balzan0,Balzan} employed proton \RR{MRI} to study random packing of monodisperse spheres.
The authors reconstructed the complete packing structure of 20,000 identical solid polymer spheres
embedded in a 0.5 L container \cite{Balzan}. The polyoxymethylene plastic spheres had
nominal diameters of 3 mm, they do not generate a signal in the \RR{MRI} scanner. These spheres were immersed
in a CuSO$_4$-doped aqueous solution. The water signal was detected, i. e. a negative of the granular
particles was obtained. The role of the CuSO$_4$ dopant is to reduce the relaxation times, $T_1$ was
lowered to about 15 ms.

The authors performed gradient echo imaging with a 7 T human scanner (Siemens) to acquire the tomograms.
The repetition time
between the scans was $T_R=5$~ms, the images were recorded with a spatial resolution of 0.2~mm.
The positions of all spheres were obtained by image analysis. The method was tested by recording the same
packing configuration twice and by comparing the number and positions of detected particles.
The authors were concerned in particular not to miss not a single particle, which would have considerable
consequences for the computed packing density.
The study demonstrated the high quality of the images and verified the reliability of the analysis.
On the other hand, the total acquisition time of the raw data was 6~h, which probably limits the
application of the method. Figure \ref{fig:balzan} shows the container filled with the embedding
liquid and the reconstruction of the bead arrangement.

Monodisperse spheres in a Taylor-Couette geometry were investigated by Moucheront et al. \cite{Moucheront2010}.
The aim of their study was the characterization of particle arrangements under shear. Their experimental setup had  19.5 cm diameter and a total height of 28 cm, fitting into a Bruker Biospec 24/80MRI scanner with 40~cm bore. The scanner operated at moderate 0.5~T.
The particles used were mustard seeds, which contain oil and can be imaged by MR. Relaxation times were
$T_1 \approx 100$~ms and $T_2\approx 40$~ms. The authors selected slices with 5 mm thickness perpendicular to the rotation axis $z$, where they achieved an in-plane spatial resolution of 0.2~mm $\times$ 0.2~mm, much better than the particle diameters of $\approx 1.5$~mm. Irrespective of the limited resolution in $z$ direction,
the authors were able to resolve the individual particles, and to identify ordering of the spheres in lanes
after sufficiently long shearing (200 rotations of the inner cylinder). In addition, the authors reported
dynamic measurements, as mentioned in Sec.~\ref{sec:dyn}.

When one is interested only in the packing density and not in the structure and arrangement of individual grains, MRI can provide 3D spatially resolved information as well. A typical application is the study of Reynolds' dilatancy of sheared grains. In a so-called split-bottom geometry (cylindrical container with
a rotating bottom disk) one can continuously shear grains and observe the formation of shear bands. Sakaie et al.~\cite{Sakaie2008} sheared poppy seeds and retrieved the packing fraction across the sheared zone  from the intensity of the \RR{NMR} signal. The resolution of the images (1.56$\times 1.56$~mm$^2$) was
not sufficient to resolve the submillimeter seeds, but the local signal proportional to the packing fraction allowed to map the dilation and shear zone profiles. During the MR scans, the rotation of the
bottom disk was stopped.

Another experiment that probed the packing fraction was performed by Kiesgen de Richter et al.~\cite{deRichter2015}. The authors were interested in the dynamics of compaction during vertical vibration of an ensemble of monodisperse grains. For that purpose, they immersed the NMR-inactive submillimeter glass beads in water and determined the local concentration of water-filled voids between the spheres. Here, only one (vertical) spatial dimension was resolved, which makes the MRI experiment quick and suppresses motion artifacts. The evolution of the vertical profile of the packing fraction was determined with a time-resolution of 5 s (the shaking frequency was in the 50 Hz range).
In a similar fashion, one can monitor the sedimentation of colloids by recording the intensity of the MR
signal of the dispersing liquid.
The sedimentation of submillimeter polystyrene spheres (NMR inactive) in a mixture of
of polyalkylene glycols was studied quantitatively with MRI by Turney et al.~\cite{Turney1995}. The same authors measured
sedimentation of rodlike particles with this technique \cite{Turney1995b}. In both cases, no resolution on the individual particles was needed, an average intensity profile was evaluated.

\subsection{Mixing and segregation}

An interesting problem in mixtures of particles, with enormous practical relevance, is mixing and segregation. These processes often take place inside the bulk of the granular bed, invisible from outside.
MRI can provide snapshots of the segregation or mixing processes. Since this technique is usually slow, one has to stop agitation  (shearing, shaking, pouring) of the sample during data acquisition and take instant images of the grains at rest.

 \begin{figure}[htbp]
\includegraphics[width=0.9\columnwidth]{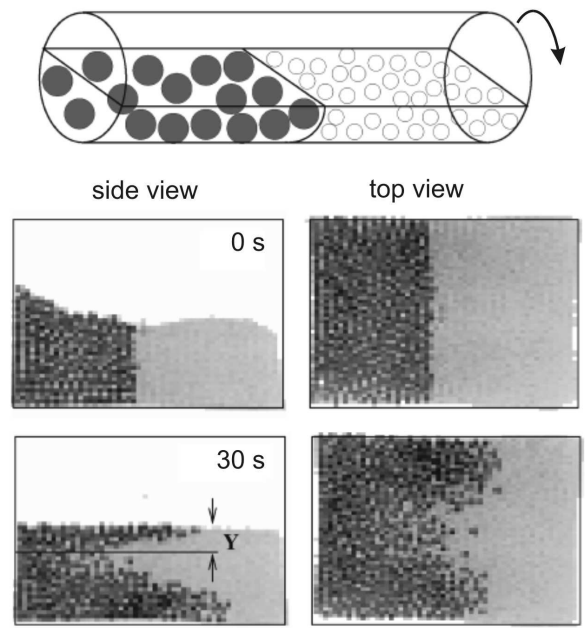}
	\caption{Sketch of a cylindrical mixer initially filled with two segments
of completely de-mixed large and small spheres, respectively (top),
and MR images of cross sections of the front between the
segments before rotation and after 30 sec of rotation with 11.4
rmp (bottom). Figure adapted from Ref.~\cite{Ristow1999} with permission,
courtesy of G. Ristow.}
\label{fig:ristow}
\end{figure}

One of the standard mixing/segregation experiments is the horizontally rotating cylinder.
Hill et al.~\cite{Hill1997} studied the size-segregation of a bidisperse mixture of spherical grains
in a half-filled horizontal cylindrical mixer with MRI. Two species,
 an \RR{NMR}-sensitive component (pharmaceutical pills of density 1.1 gcm$^{-3}$ with 1~mm diameter)
and an \RR{NMR}-insensitive component (3~mm plastic spheres of density 1.6 gcm$^{-3}$) were initially mixed in
50\%:50\% volume ratio.
The mixer was an acrylic cylinder with 75 mm diameter, additional pharmaceutical spheres were taped at the outside as reference points.
The scanner with a 31~cm horizontal bore (Nalorac Cryogenics Corp.) worked at 1.9 Tesla.
The tomograms mapped the distribution of the \RR{NMR}-active species.
Even though the resolution was not sufficient to identify the individual grains, segregated clusters could be clearly detected and imaged.
The main result of this study was that the constitution of the surface of the granular
bed, i. e. the profile of the surface and the distribution of small and large beads in the
flowing layer, does not give the full information
necessary to understand the formation of segregation stripes. The small species forms an axial core
inside the granular bed, and the radial instability of this core of small grains leads to an axial modulation that precedes and probably triggers the subsequent axial segregation at the surface  \cite{Hill1997,Hill1997a}.

The same hardware configuration was used by Ristow and Nakagawa \cite{Ristow1999} to map the front of two segments
of granulate in a similar mixer of 10 cm length and 7 cm diameter. Initially, a completely segregated state of small ($\diameter$ 1 mm) and large ($\diameter$ 4 mm) pharmaceutical pills was prepared
(Fig.~\ref{fig:ristow}, top). The particles contained a vitamin oil and are thus \RR{NMR}-sensitive.
After rotating the cylinder with 11.4 rpm, snapshots were taken in intervals of 15 sec
($\approx 2.85$ rotations). Figure \ref{fig:ristow} shows slices of the granulate
in the beginning and after 5.7 rotations. The image resolution was $64\times 64\times 64$ pixels. Even though the individual particles were not resolvable (pixel dimensions larger than 1~mm), one can clearly
identify the two species in the tomograms, and see the evolution of the front.

\begin{figure}[htbp]
\center
\includegraphics[width=\columnwidth]{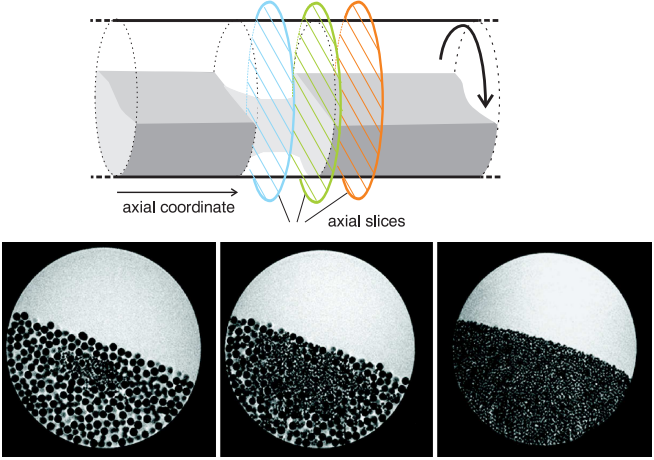}
	\caption{Horizontal cylindrical mixer with glass spheres (1.5 mm and 0.55 mm diameter)
embedded in water. Initially, two axial bands with small beads were prepared, separated by a band
with large beads. During rotation, an axial core of small beads forms, through which the particles are transported. This finally causes extinction of one of the bands.
The top image sketches the positions of three selected slices from the 3D tomogram that are shown in the
bottom row.}
\label{fig:finger}
\end{figure}

A similar experiment was performed by Finger et al. \cite{Finger} with the aim to reveal the mechanism
of coarsening of axial segregation band structures. The cylindrical mixer was half-filled with a
bidisperse mixture of glass beads with 1.5~mm and 0.55~mm diameter in 50\%:50\% volume ratio,
then it was filled up with water. In these experiments, a Bruker BioSpec 47/20 MRI scanner
(4.7 T, 200 MHz proton resonance) was used. The MR images show the water distribution, the beads appear black.
The spatial resolution of 0.5 mm allowed to resolve the individual particles in the mixtures.
Fig.~\ref{fig:finger} shows cross sections of the 36.8 mm diameter mixer at three different locations,
extracted from the 3D tomogram.
The small grains at the initially flat interfaces have started to penetrate into the
region filled with large particles and formed an axial core, similar to Fig. \ref{fig:ristow}. By mapping the structure of this core in a sequence of MR images, the particle exchange between
neighboring bands of segregated particles was monitored.

\begin{figure}[htbp]
\includegraphics[width=0.8\columnwidth]{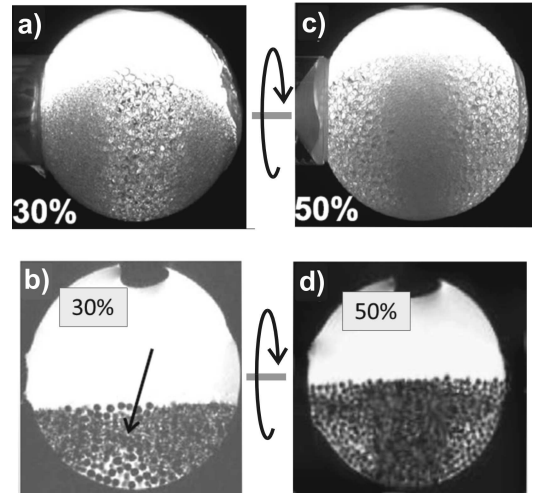}
	\caption{Spherical mixer rotated about a central horizontal axis (perpendicular to the viewing direction).
Initially, the bidisperse grains were well mixed. After $\approx 100$ rotations, the material is completely segregated. Optical images a and b show the surface segregation, while MRI (c,d show vertical slices through the mixer center) reveals the core structures. The dark spot on the top of the mixer is a trapped air bubble.}
\label{fig:naji}
\end{figure}

Figure \ref{fig:naji} presents MR images with the same hardware, same material, but in a spherical mixer
geometry \cite{Naji}, a and b are optical images which show only the surface distribution of the small
(dark in the image) and large (brighter in the image) glass beads.
Images c and d are vertical slices from the center of the 3D tomograms. The MR images show the water in the mixer, the beads appear black.

The transport of \RR{NMR}-active mustard seeds mixed with glass spheres or sugar balls in the horizontal rotating cylinder geometry was investigated by Metcalfe et al.~\cite{Metcalfe1996,Metcalfe1999}.
By combining different materials in bidisperse mixtures, they were able to distinguish the influences
of size and mass density on radial and axial segregation.

Sederman et al.~\cite{Sederman2007} measured the segregation of millet and poppy seeds in the rotating
mixer. They used a Bruker Biospec AV (85 MHz, 2 T). The millet seeds with 2.5 mm diameter are larger than the poppy seeds ($1\times 0.7 \times 0.7$ mm$^3$), they segregate radially at the outside, while the poppy seeds form an inner core. The images are not as easily interpreted as the previous experiments, since both components, millet and poppy seeds, contribute to the \RR{NMR} signal. However, the relative signal intensity of the poppy seeds is 10 times larger than that of the millet seeds so that the contrast is sufficient.

\begin{figure}[htbp]
\includegraphics[width=\columnwidth]{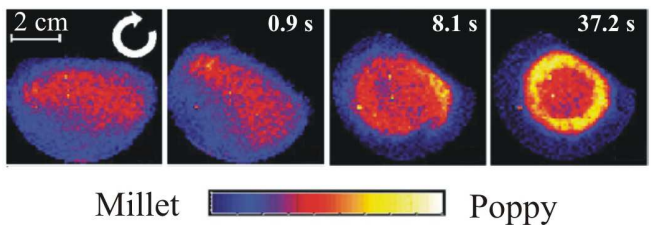}
	\caption{2D FLASH images of a cross section of a cylindrical mixer containing millet and poppy seeds, during rotation at 10 rpm. The first image was taken before the rotation started, times are measured with respect to the start of the rotation. The poppy seeds give a much stronger NMR signal. A radial segregation takes place within a few seconds, the smaller poppy seeds form an axial core that is surrounded by the millet
seeds. 0.9 s correspond to 54$^\circ$ rotation angle, 37.2 s to 6.2 full revolutions. Image adapted from Ref.~\cite{Nguyen2011} with permission, courtesy of L. F. Gladden.}
\label{fig:nguyen}
\end{figure}

Substantial progress in the dynamic study of the same system was achieved by Nguyen et al.~\cite{Nguyen2011}.
High-resolution tomograms were taken with the RARE technique \cite{Henning} while the mixer was at rest, as in the above mentioned study. But moreover, an ultrafast MRI technique, the 2D FLASH sequence (Sec.~\ref{sec:FLASH}), allowed to measure images of 2D slices
while the cylinder was rotating at 10 rpm. Images were obtained in 0.3 s intervals. Figure
\ref{fig:nguyen} shows exemplarily some cross sections of the mixer during the formation of the
radial segregation pattern.

There is a third option to obtain MR images of granular materials in mixing and segregation processes.
One can label individual beads with a liquid. Such an approach was employed by Sommier et al.~\cite{Sommier2001}, using 1 mm diameter sugar beads that were either left untreated (NMR inactive) or soaked into silicon oil (NMR active). The authors filled a mixer with well separated regions (50 \%: 50 \%) of both grain types and recorded
the MR images (resolution $\approx 0.95 \times 0.95$~mm$^2$ in slices separated by 3.5 mm, $64 \times 64 \times 64$ voxels) after certain intervals of rotations. From the analysis of the distribution of the NMR sensitive grains in the container, the mixing characteristics of the device were documented. The authors compared this mixing
characteristics with the segregation of poppy seeds and sugar beads of different size in the same
mixer, starting with a well-mixed state. A slightly extended study of the same system was published by
Porion et al. \cite{Porion2004}.

\subsection{Displacement, diffusion, and flow}
\label{sec:dyn}

The simplest way to detect displacements in granular material is the use of labeled grains, viz. \RR{NMR}-active probe grains embedded in non-active material, or grains that can be distinguished from their
surroundings by size or shape. MR images are
taken before and after some agitation of the material, such as shearing or mixing.
In the above described split-bottom geometry (rotating disk at the bottom of a cylindrical container)
the 3D flow profile and the geometry of the localized shear zone was measured with MRI of tracers
by Cheng et al.~\cite{Cheng2006}. The sample consisted of a mixture of rajagara (amaranth) seeds with 5 \% of poppy seeds. Both seeds have comparable diameters and densities. The NMR signal is generated mainly by the poppy seeds that serve as markers. By comparing MR images recorded after successive rotation intervals, the flow field was reconstructed from cross-correlations of these images.

\begin{figure}[htbp]
\includegraphics[width=\columnwidth]{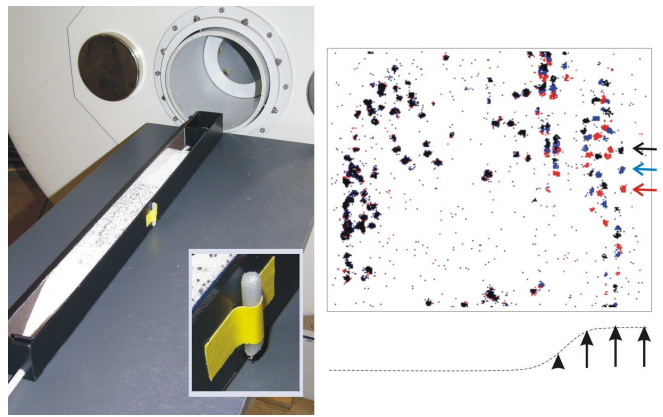}
	\caption{U-shaped rail consisting of two L-shaped sliders and the Bruker MRI scanner, the inset shows
a test tube filled with liquid to mark the slider displacement. On the right hand side, the overlay of the same
slice from three successive MR images is shown, the poppy seed tracers in the three images are marked in red, blue and black.
Their displacement is used to construct the flow field (sketched qualitatively below the tomogram).}
\label{fig:tamas}
\end{figure}

Another simple shear experiment with \RR{NMR}-active probes is shown in Fig. \ref{fig:tamas}.
The U-shaped horizontal shear cell consisting of two parallel L-shaped sliders is filled with corundum particles (left hand side of the cell) and glass spheres (right hand side of the cell). Shifting of one of the sliders leads to the formation of a shear zone, the grains at one side of the shear zone remain in their original position with the slider at rest, the particles on the other side are displaced.
The shear zone is shifted and bent in such a stratified granular bed. For a quantitative investigation,
one can map the displacement and the width of the shear zone at the surface with color-marked grains.
The subsurface shear zone profile has to be determined destructively by excavation of the
granular material.
MRI is a useful alternative tool to measure the displacement field and to map the shear zone profile,
non-invasively. For that purpose, B\"orzs\"onyi et al. \cite{Borzsonyi2011} added poppy seeds
(size $\approx 0.75$ mm) as markers in low concentration to the MRI-invisible glass beads and corundum
grains.
With a Bruker BioSpec 47/20 MRI scanner (200 MHz proton resonance frequency, 4.7 T), a resolution of
0.156~mm per pixel was achieved in each slice. Typically, 20-25 slices were recorded in 1~mm distance.
Figure \ref{fig:tamas}, left, shows the U-shaped cell consisting of two sliders.
A small water-filled tube attached to one slider (see inset) served as displacement marker.
MR images were recorded at slider displacements in 2.5 mm steps, and the tracer positions were extracted.
Arrows at the side of the tomogram slice in Fig. \ref{fig:tamas} mark exemplarily three subsequent
positions of one of the tracer particles. From the displacements of all tracers, a
shear zone profile was reconstructed in each slice. Qualitatively, this is sketched by the arrows below the tomogram.
Such tracer experiments have to be performed and evaluated with caution, one has to make sure that the tracers do not demix and segregate if they differ in size or weight from the bulk material.
In the described experiment, the total shear was small so that such demixing was negligible.

Measurements of diffusive axial transport of grains in a rotating cylindrical mixer
were reported qualitatively by Nakagawa et al. \cite{Nakagawa1997b} and later quantitatively by
Fischer et al. \cite{Fischer2009}. The problem was posed by
Khan et al. \cite{Khan2005}, who had described an optical experiment with color-marked glass beads:
They concluded from their observations that the axial transport of the grains is subdiffusive.
Their optical method had the problem that it provided only an
integral signal of black tracer beads embedded in a translucent surrounding. With MRI, each individual particle ($\diameter$ 4 mm and 2 mm glass beads) in water could be resolved and their distributions after  well-defined numbers of revolutions could be extracted \cite{Fischer2009}. In these experiments, only the particle distribution density was evaluated, individual particle trajectories were not required. MR images were taken while the mixer tube was stopped, every 2.5 rotations. From the distribution of initially prepared configurations (e. g. a thin slice of large grains embedded in small ones) it was established
that the small beads in this bidisperse system undergo normal diffusion along the mixer axis when the
mixer tube  rotates. These MRI results supported results of discrete elements methods simulations
\cite{Taberlet2006}.

\begin{figure}[htbp]
\includegraphics[width=\columnwidth]{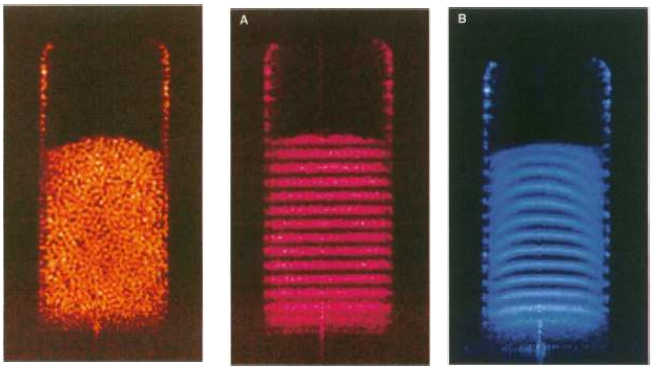}
	\caption{MR images of a cylindrical container ($\diameter 16.5$ mm) filled with poppy seeds.
Left: Image of 1 mm thick slice of the MR tomogram. Middle: 2 mm slice of magnetization labeled
grains in the same container, $M_0$ was sinusoidally modulated along the vertical axis by the preparation sequence,
Right: same after one vertical shake of the container. The particles have shifted by convective flow, thus the
slices appear distorted.
Reproduced with permission from Science 267, 1632 (1995). Copyright 1995 AAAS.}
\label{fig:ehrichs}
\end{figure}

These MRI techniques, which rely on the detection of individual particles, are certainly advantageous over invasive characterization methods (excavation),
but in general, they can hardly compete with X-ray CT, which can often be performed faster and
with higher resolution, and the choice of materials is not as limited as with MRI.
However, there are MR imaging techniques that provide additional opportunities.
Labeling spins in certain regions of the sample is an appropriate tool to detect re-arrangements of particles, for example after shaking a container, or after rotation of a mixer.
This approach has been successfully introduced in granular matter research by Ehrichs et al.~\cite{Ehrichs1995} who studied the displacement of particles in a container after one single shake. The authors were interested in the detection of convection flow. For that purpose, the cylindrical container ($\diameter 16.5$ mm) was filled with poppy seeds.
Figure~\ref{fig:ehrichs}, left, shows one slice of the \RR{MR} tomogram (1 mm thick) in the container before shaking, where the individual poppy seeds are resolved. The central image shows a thicker slice
(so that individual poppy seeds are no longer distinguishable) in the same container. Layers were selected by spin tagging with an appropriate preparation sequence of pulses and gradient
(see Sec. \ref{sec:velocimetry}).
In the right hand image, the same container
is shown after one tap between preparation sequence and signal acquisition. One can easily recognize details of the grain rearrangements: The material is compactified, the material mixes at the bottom and side walls, and there is an obvious downward displacement of the grains near the container walls with respect to those near the center of the container in the same layer.
This displacement is the signature of an earlier described convection roll \cite{Knight1993} with up-flow near the central axis and down-flow near the
outer walls.
Technical details \cite{Kuperman1995}, and further results of the granular experiment, e. g. observations with tagged vertical slices \cite{Knight1996}, were also published.

Another option to obtain flow profiles of NMR-active grains or of interstitial fluids in a granular
bed are gradient echo experiments. As mentioned above, the application of two field
gradients of opposite sign but otherwise same strengths and durations refocuses the magnetization
in absence of translational motion. If the spins diffuse during the evolution time between the two gradients, the echo signal is attenuated. In case of a drift of the spins in the direction of the gradient, the echo obtains a phase shift that depends linearly upon the flow velocity component along
the gradient direction. The signal will then contain the velocity information of the excited spins, it is encoded in the phase of the signal. The spin density has to be determined separately (without gradients).
Even though this method yields more detailed information than the spin tagging approach, it is not
always advantageous. The evaluation of the phase shifts is less straightforward than the
spatial intensity profile in a spin-tagging experiment.

Both methods can of course be combined to obtain a more comprehensive view of flow in complex geometries.
Such a strategy was chosen, e.g. in a study of flow of a fluid (water) through a cylindrical tube ($\diameter$ 19 mm) filled with randomly packed glass beads ($\diameter$ 4 mm) and different catalyst pellets \cite{Ren2005}. The arrangement of the solid particles was determined from the images of the fluid at rest in several 2D cross-sectional slices. Then, a spin-tagged layer of the liquid was selected with a $\pi/2$ pulse during a magnetic field gradient along the tube axis. The displacement of the liquid was measured as a function of the evolution time. In addition, a flow-encoding technique was used to encode the local velocity of the fluid in the phase shift of the echo signal. The authors obtained a 1D radial profile of the velocity of the water streaming through the pipe as well as the 2D spatially resolved mean flow in selected slices. Of course, there are also numerous
experiments with dense suspensions that make use of similar techniques (see, for example,
Refs. \cite{Altobelli1991b,Graham1991,Abbott1991,Altobelli1997}).

\begin{figure}[htbp]
\includegraphics[width=\columnwidth]{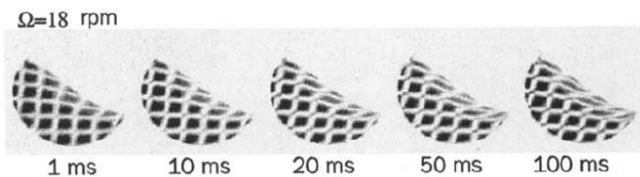}
	\caption{Cross section of a horizontal rotating cylindrical container, half-filled
with mustard seeds. The images show spin-tagged slices recorded during the rotation at 18 rpm.
During 100 ms, the rotation angle is approximately 10$^\circ$.
Reproduced with permission from Exp. Fluids 16 54 (1993). Copyright 1993 Springer Verlag.}
\label{fig:nakagawa1993}
\end{figure}

Flow profiles of granular matter or slurries have been studied by MRI in a number of other geometries
(with either spin tagging or phase encoding or both),
such as in Couette cylinders \cite{Mueth2000,Moucheront2010}, in spouted beds \cite{Kawaguchi2007}, and in horizontal cylindrical mixers \cite{Altobelli1993,Yamane1998,Nakagawa1993,Nakagawa1997,Seymour2000}.
Flow of seeds in a rotating horizontal tube was mapped by Nakagawa et al.~\cite{Nakagawa1993,Nakagawa1997} with two different methods: Spin tagging allows to select not only a striped 1D periodic pattern (see above),
one can also create a rectangular grid pattern. When the mixer tube rotates, part of the grid pattern
will rotate with the container in a solid-like motion of the granulate, but in addition, rearrangements
of individual grains will distort the pattern as seen in Fig.~\ref{fig:nakagawa1993}. This can be
compared to simulations \cite{Yamane1998} to test numerical models. In the same system, Nakagawa et al.~\cite{Nakagawa1993,Nakagawa1997} measured the flow velocities directly, by flow-encoding the
phase of the MR signal.

Among the problems with high practical relevance in granular physics is the discharge of hoppers.
Inside 2D hoppers with transparent walls, the flow field can be mapped easily with colored markers.
In 3D, non-invasive imaging techniques are inevitable. Gentzler et al.~\cite{Gentzler2009} used a
pulsed gradient echo technique to determine the flow in a vertical slice.
First, the spin density was measured, then, a fixed vertical gradient pulse was applied which encodes the vertical particle displacement in a phase difference. The result was a 2D map of the vertical flow
component in a central slice of the hopper.
An alternative method to visualize the flow profile here could be spin tagging as well.

A particular challenge is dynamic imaging of vibrated systems during the periodic excitation
\cite{Caprihan1997,Huan2004,Huntley2014}. This can been
achieved by triggering the MRI data acquisition synchronously with the vibrator. Such an experiment was described by Caprihan et al.~\cite{Caprihan1997}. Single line scans with
echo times of 3.5 ms allowed to image slices during vibrations with 24 Hz frequency.
Even though individual particles (2 mm oil filled plastic capsules) were not resolved,
the authors could nicely visualize the dynamics of so-called arches that were formed in the vibrated granular bed.

\subsection{Dynamic measurements}

Even more demanding is the fast monitoring of single dynamic events.
An example is the dynamics in gas-fluidized granular beds, which has particular importance in fluidized bed reactors. Such a reactor contains a deep layer of granular material into which a high-velocity gas stream is blown from below.
One of the technologically interesting questions is the mixing of
a pulse of particles added to a fluidized bed of other particles. This was studied by Fennell et al. \cite{Fennell2005} by means of MRI. With the aid
of the above mentioned FLASH sequence, a vertical 1D profile of such an air-fluidized granular bed was imaged with a time resolution of 12 ms and a spatial resolution 1/1.6 mm. Thereby, poppy seeds served
as MR tracers. The signal intensity mapped the distribution density of the seeds and revealed their penetration into the fluidized bed of slightly smaller sugar balls.

\begin{figure}[htbp]
\includegraphics[width=0.5\columnwidth]{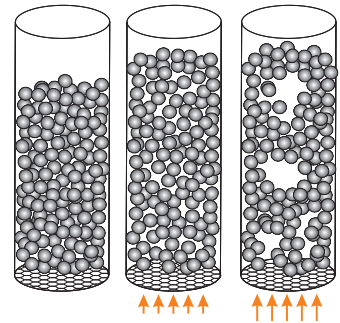}
\caption{
Sketch of an air-fluidized granular layer in three phases;
left: no flow or moderate flow with compact bed;
middle: stronger air flow with uniform dilution;
right: strong flow with bubbling and jetting.}
\label{fig:airflu}
\end{figure}

A related topic is the dynamics of jet and bubble formation in such gas-fluidized beds,
as sketched in Fig. \ref{fig:airflu}. The gas flow velocity determines the structure and dynamics of the granular bed. This process has been studied with various imaging techniques such as X-ray tomography \cite{Bieberle2010}, MRI
\cite{Harms2006,Rees2006,Mueller2006,Mueller2007,Mueller2008,Mueller2011,Koehl2013,Koehl2014,Zhang2015,Penn2016}
or electric capacitance imaging \cite{Makkawi2002,Chandrasekera2012}.
Gas and particle motion in the reactor represent a complex multiphase-flow problem.

A technique to obtain a spatially resolved velocity profile in such a bed by \RR{MRI} was proposed by Harms et al.~\cite{Harms2006}. In contrast to the above mentioned flow profile studies in a compact bed \cite{Ren2005}, the grains reorganize dynamically during the air flush, so fast data acquisition is
mandatory.
Rees et al. \cite{Rees2006} managed to take snapshots of slices
in different heights of an air fluidized granular bed of (unspecified) seeds. Packing densities in 41 horizontal slices (1~mm thick) were obtained during an acquisition time of 730~ms, with an echo time of 2.51 ms. The in-plane spatial resolution was 128~$\times 128$~pixels in a 55~mm $\times 55$~mm field of view. In the same study, the grain dynamics was mapped by phase-difference velocity encoding, using a slice-selective spinecho sequence. The lateral spatial resolution was the same, the slice thickness 5~mm.
The dynamics of the formation and collapse of the bubble- and channel-like voids in the reactor was measured by M\"uller et al. \cite{Mueller2006} with
a Bruker DMX 200 spectrometer ($200$~MHz proton frequency) and gradients up to 0.139 T/m.
The standard 1D FLASH sequence allowed to record profiles with 1.25~mm pixel resolution in 384~$\mu$s acquisition time, and higher 0.625 mm spatial resolution with 853 $\mu$s acquisition time.
This spatial and temporal
resolution was sufficient to follow the dynamics of individual gas bubbles in the granular bed.
Recent developments allowed to record 2D MR image sequences with a temporal resolution of 26 ms/frame \cite{Penn2016} at a spatial resolution of 3.1~mm $\times 3.1$~mm. The imaging technique was improved by (I) performing parallel MRI with tailored multi-channel receiver hardware, (II) by employing fast gradient readouts for time-efficient data sampling, and (III) by studying special engineered grains that contain a high proton density liquid (oil encapsuled in 1~mm spherical rigid agar shells).

\begin{figure}[htbp]
\center
 \includegraphics[width=\columnwidth]{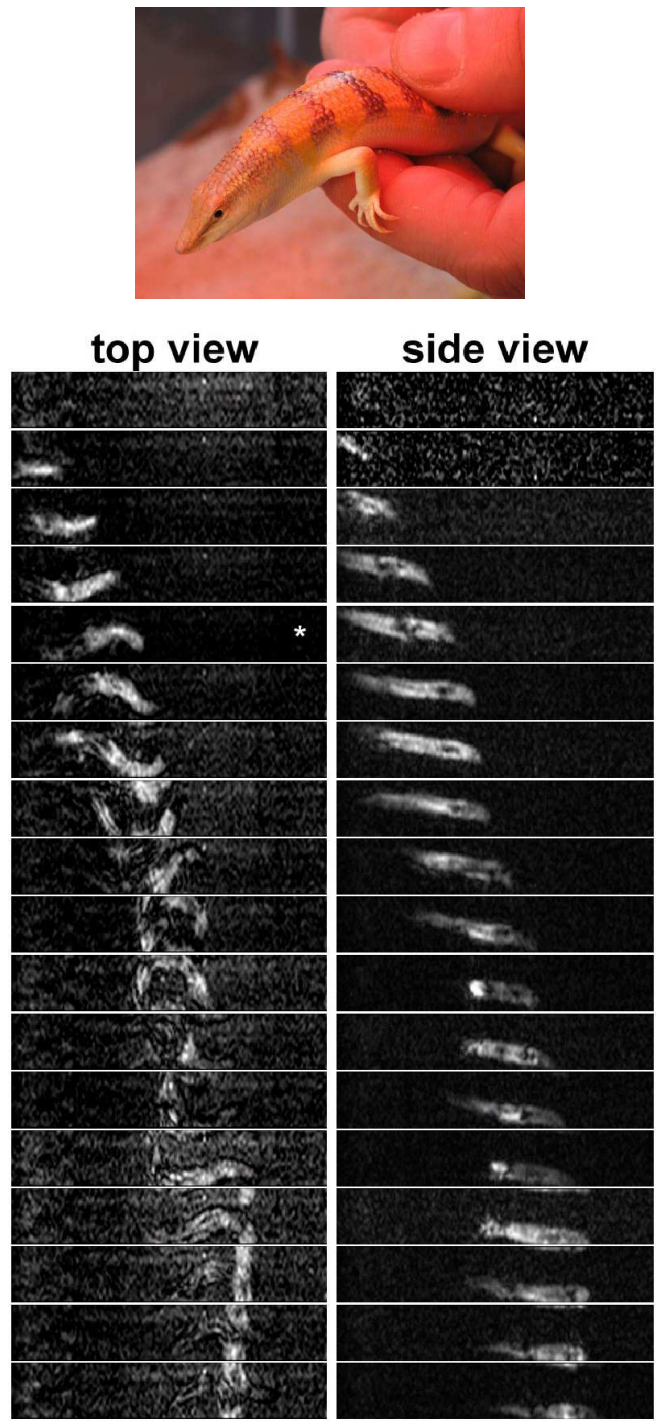}
	\caption{ Top: Scincus scincus. Bottom:
MR image series of the burying process of a sandfish ({\em Scincus scincus}).
In the top view, one recognizes a sinusoidal (snake-like) movement, the
side view at the same time shows no significant contortion of the body.
The time interval between the images is 120~ms. Due to the \RR{MRI} protocol, parts of the object
leaving the imaging area at one edge reappear at the opposite edge.
Images reprinted from Ref.~\cite{Baumgartner2008}, courtesy of W. Baumgartner.}
\label{fig:baumgartner}
\end{figure}

Finally, we describe a biological application which has very close connections to granular matter physics. The sandfish ({\em Scincus scincus}) is a lizard that can dive very quickly into dry sand to escape predators. Submerged, it performs swimming-like motions in the sand bed, and it can traverse considerable distances very fast. The understanding of the mechanical aspects of this subsurface locomotion is of course of great technical interest. In order to understand the details of the
animal's exercise techniques, it is necessary to scan inside the sand bed.
This can be achieved by means of
X-ray tomography \cite{Maladen2009,Maladen2011}. The first tomogram of the sandfish swimming in a
sand bed was, however, obtained by Magnetic Resonance Imaging. Baumgartner et al.~\cite{Baumgartner2008}
placed a cylindrical box filled with sand, 40 cm long, and 22 cm in diameter, inside the birdcage head
coil \RR{(two concentric end rings connected by parallel straight elements, so-called rungs)}
of a 1.5 T whole-body MRI scanner ({\em Siemens Magnetom VISION}).
The \RR{NMR} signal comes solely from the animal, the dry sand does not contribute to \RR{that} signal. Two projections, dorsal and lateral side views were recorded at an image rate of 120 ms, with spatial resolution of 1.56~mm $\times 1.17$~mm. In these measurements, a modified FLASH sequence was employed
(flip angle 8$^\circ$), 5 ms repetition rate, with two echo readouts. No slice selection gradient was employed, so that the images are simple projections of the spin density (similar to X-ray transmission images).
With the read encoding gradient in axial direction and the phase encoding gradients for the first echo
in coronal projection, for the second echo in transverse direction, the two projections of the tomogram
(top and side views) could be recorded simultaneously, 24 $\times$ 256 pixels spatial resolution required the 120~ms acquisition time for each image. The result is seen in Fig.~\ref{fig:baumgartner}.
The MRI data are comparable to X-ray CT results of the same animal experiment \cite{Maladen2009,Maladen2011}. In both the \RR{MRI} and X-ray studies, only the projection(s) of the moving
animal were recorded, with somewhat better resolution in the X-ray images. In the latter, small markers were fixed to the animal to follow the details of the limb motions. Probably, the advantage of MRI is
that the size of the experiment box is limited only by the scanner dimensions. In the X-ray study,
a shallow 10 cm deep bed was used. Thicker layers of sand will deteriorate the image.
In this experiment, the potential of MRI to select only one slice was not required,
but in principle this could have been achieved by appropriate slice selection during the $\pi/2$ pulses.

\section{Comparison with competing techniques}

The attentive reader may have noticed from the selection of examples that after some vivid and promising
phase with rapid progress in the 90ies of last century, MRI activities in the field of granular matter research gradually ceased, or at least stagnated.
This is fact even though MRI in medical applications is well established and still rapidly developing, and even though granular matter physics remains to be a hot topic with many
unsolved fundamental questions. There may be several reasons for this trend. Among the major factors are
probably the immense costs for MRI scanners, which hardly give adequate return in pure granular matter
research. Moreover, X-ray CT as a strong competitor has some obvious advantages over
MRI: The expenditure of time for a complete 3D tomogram is generally much higher in MRI.
There is only a limited selection of materials (liquids or liquid-containing particles) suitable for
MR Imaging. The majority of MRI experiments in particle imaging have been performed with seeds and pharmaceutic pills. This limits the choice and combination of grain sizes and shapes considerably.
X-ray CT is clearly less selective in that respect.

Because of the strong magnetic fields necessary for MR scanners, not only the samples but also the immediately attached parts of the setup must be free from
ferromagnetic elements. This is particularly restrictive when real-time data are to be sampled in
a mechanical device. If a sample is to be agitated mechanically in the scanner, all mechanical parts
must be free from iron components. Motors or shakers must be driven by air pressure or
placed in sufficiently large distance from the scanner. Even non-magnetic metal parts can disturb
with eddy currents. Here, X-ray CT is much less vulnerable as well.

Another limitation that can be more restrictive than in X-ray CT is the available space.
In MRI, the largest commercial wide-bore scanners offer some 70 cm bore width, which may be tolerable
in most applications, but those scanners are not typically available.
Certainly, some technical aspects are also impractical in typical research
environments: The scanner cannot be switched off. Unlike in X-ray CT, most preparatory steps
must be performed outside the scanner premises. This is particularly unfavorable in granular matter
since any transport of specially arranged samples can cause tremors and disturb the prepared states.

Nevertheless, a clear potential of Nuclear Magnetic Resonance in particle imaging lies in the highly
sophisticated techniques that have been developed for several special purposes. If one is interested
not in a full 3D image of a sample but in the image of an arbitrary 2D slice inside, an enormous amount
of time can be saved. There are no adequate techniques in X-ray CT that can save data acquisition time
by selecting only a part of the sample volume. With appropriate pulse sequences, such 2D images can be
obtained in milliseconds.
An advantage of MRI is the option to selectively label slice arrays or grids of particles in a
granular bed, to follow the displacement of such labeled regions in the sample, the rearrangement
of grain ensembles and flow or diffusion effects.

Another option that is not available with X-ray is flow encoding, the reconstructed image contains information about dynamic processes in the sample. Flow encoding techniques can reveal the local motion
of grains, even microscopic displacements, without the necessity to track individual tracers.

Even though X-ray CT is by far the strongest competitor, there are yet other imaging techniques useful
for the study of granular materials. Positron Emission Tomography (PET) may provide information on segregation phenomena or other structural parameters of granular beds (see, e. g. \cite{Windows-Yule2016}). In comparison with MRI, the PET sample preparation is more demanding,
and PET also suffers from a limited selection of materials.
Moreover, the latter method relies on ergodicity of the sample, i. e. the statistical properties of the trajectories of probe particles must agree with the ensemble averages.
PET samples have a limited lifetime, whereas an MRI sample can be re-used practically
without limitations.

Optical scanning of ensembles of transparent particles in a (fluorescent) immersion fluids with a
Laser line or Laser sheet can provide 3D images of particle arrangements as well. This method is much cheaper and it often can be incorporated in existing experimental setups with moderate efforts. It can
be combined with particle tracking and measurements of granular flow profiles. However, this method has serious restrictions \cite{Dijksman2012}. The choice of materials, in particular the combination
of transparent particles and an appropriate immersion fluid, is substantially more problematic and
restrictive than in MRI.

Other imaging techniques like electrical capacitance tomography \cite{Huang1988}
have been proposed occasionally to study granular matter \cite{Makkawi2002,Chandrasekera2012}.
The latter is not tomography in the genuine sense. It uses multiple electrodes to produce
very low resolution images. The advantage of this method is that it is fast, and inexpensive.

Summarizing, here are some figures of merit of MRI:
\begin{itemize}
\item
the time resolution for fast 2D imaging is in the range of 100 ms and even slightly below, and much faster if only 1D profiles are recorded,
\item the spatial resolution is in the submillimeter range,
\item spatial limitations are $\approx 40$~cm $\times 40$~cm, in wide-bore scanners even larger.
\end{itemize}
Advantages over competing techniques are
\begin{itemize}
\item low energy, no radiation damage, the method is harmless for experimenters,
\item scanners are widely available in medical centers,
\item no special sample preparation is needed,
\item spatially resolved flow and diffusion measurements are possible,
\item a differentiation of materials is possible by selective labeling.
\end{itemize}
The user pays for these advantages with
\begin{itemize}
\item expensive hardware, expensive maintenance,
\item only limited choice of materials,
\item the necessity to have a metal-free experimental setup,
\item often very long acquisition times,
\item susceptibility generated distortions and other image distortions,
\item reconstruction artifacts particularly in fast imaging techniques.
\end{itemize}

It goes without saying that all non-invasive screening techniques like X-ray CT and MRI
are advantageous over destructive methods such as excavation or
mechanical disassembly of packings. In that respect, these tomographic techniques have opened completely new perspectives in granular matter research. The potential for imaging of fast
dynamic processes and the uncomplicated, quick and efficient 2D imaging of selected inner slices of
a sample guarantees the competitiveness of Magnetic Resonance Imaging in many applications in the future.

\section{Acknowledgments}
The author is indebted to Oliver Speck for critical reading of the manuscript and useful suggestions,
Stephen A. Altobelli, Riccardo Balzan, Werner Baumgartner, Lynn F. Gladden, Heinrich Jaeger, and Gerald Ristow are kindly acknowledged for permissions to reproduce figures.

\bibliographystyle{unsrt}

\end{document}